\documentclass[useAMS,usenatbib]{mnras}

\usepackage[T1]{fontenc}
\usepackage{ae,aecompl}

\usepackage{mathtext,amssymb,amsmath}
\usepackage{epsfig}
\usepackage{graphics}
\usepackage{url}
\usepackage{adjustbox}
\usepackage{times}

\usepackage{graphicx}	
\usepackage{amsmath}	
\usepackage{amssymb}	
\usepackage{color}
\usepackage[english]{babel}
\usepackage{graphicx}
\usepackage{hyperref}
\usepackage{listings}
\usepackage{lscape}
\usepackage{natbib}
\usepackage{url}
\usepackage{breakurl}
\usepackage{xspace}
\bibpunct{(}{)}{;}{a}{}{,}

\newcommand{\Teff}{$T\mathrm{\hspace*{-0.4ex}_{eff}}$\,}
\newcommand{\logg}{$\log\,g$\hspace*{0.5ex}}
\newcommand{\pg}{PG\,0948+534}
\title[Spectral analysis of the extremely hot DA white dwarf \pg]
      {Spectral analysis of the extremely hot DA white dwarf \pg}

\author[K\@. Werner et al.]{
K\@. Werner,$^{1}$\thanks{E-mail: werner@astro.uni-tuebingen.de}
T\@. Rauch$^{1}$
and 
N\@. Reindl$^{2}$
\\
$^{1}$Institut f\"{u}r Astronomie und Astrophysik, Kepler Center for Astro and
Particle Physics, Universit\"{a}t T\"{u}bingen, Sand 1, 72076 T\"{u}bingen,
Germany\\
$^{2}$Department of Physics and Astronomy, 
           University of Leicester, 
           University Road, 
           Leicester LE1\,7RH, 
           UK
}

\pubyear{2018}

\begin{document}
\label{firstpage}
\pagerange{\pageref{firstpage}--\pageref{lastpage}}
\maketitle

\begin{abstract}
There is a striking paucity of hydrogen-rich (DA) white dwarfs (WDs)
relative to their hydrogen-deficient (non-DA) counterparts at the very
hot end of the WD cooling sequence. The three hottest known DAs
(surface gravity \logg $\geq$ 7.0) have effective temperatures around
\Teff = 140\,000\,K, followed by only five objects in the range
104\,000 -- 120\,000\,K. They are by far outnumbered by forty non-DAs
with \Teff = 100\,000 -- 250\,000\,K, giving a DA/non-DA ratio of
0.2. In contrast, this ratio is the inverse of that for the cooler
WDs. One reason for this discrepancy could be uncertainties in the
temperature determination of hot DAs using Balmer-line
spectroscopy. Recent investigations involving metal-ionization
balances in ultraviolet (UV) spectra indeed showed that the
temperatures of some DAs were underestimated, but the paucity of
extremely hot DAs prevailed. Here we present the results of a UV
spectral analysis of one of the three hottest DAs, \pg. We find that
its temperature was strongly overestimated by recent Balmer line
analyses. We correct it downward to 105\,000 $\pm$ 5000\,K,
aggravating the hot-DA paucity. The Balmer-line problem encountered
previously is not resolved by our non-LTE line-blanketed model
atmospheres. We speculate that it might be related to the possible
presence of a magnetosphere. This is supported by the V-band
variability that shows a period of $P=3.45$\,d (amplitude 0.19\,mag),
which we interpret as the star's rotation period. The metal abundances
in \pg\ are affected by atomic diffusion and we conclude that the
onset of diffusion in hot DAs occurs when they cool below \Teff
$\approx 105\,000$\,K. We discuss the possibility that the paucity of
very hot DAs is a consequence of their fast evolutionary rate.
\end{abstract}




\section{Introduction}
\label{sect:intro}

\begin{figure}
 \centering\includegraphics[width=1.0\columnwidth]{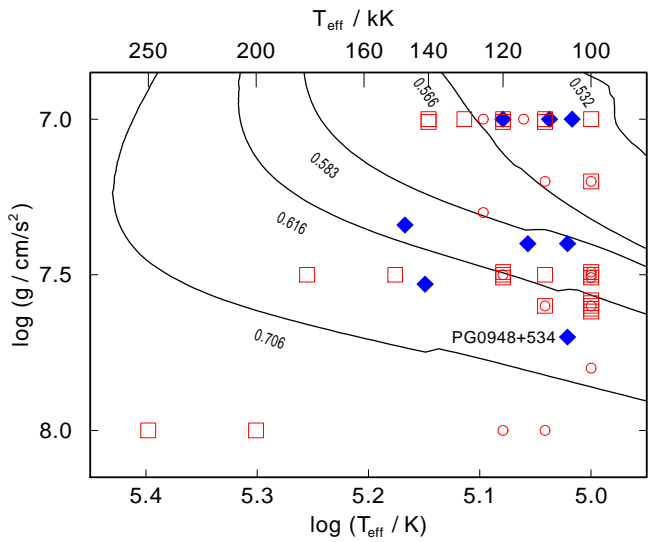}
  \caption{Presently known DA (blue rhombs) and non-DA (red symbols)
    WDs with \Teff $\geq$ 100\,000\,K and \logg $\geq$ 7.0. Red squares
    and circles denote PG1159 stars and DO WDs, respectively. The
    position of \pg\ is shown according to the result of this
    work. Black lines are evolutionary tracks of post-AGB remnants
    with masses as labeled \citep[in $M_\odot$, from][metallicity
      $Z=0.01$]{2016A&A...588A..25M}.}
\label{fig:gteff}
\end{figure}

The number of hydrogen-rich (spectral types DA and DAO) white dwarfs
(WDs) with effective temperatures ($T\mathrm{\hspace*{-0.4ex}_{eff}}$)
of 100\,000\,K or higher and \logg $\geq$ 7.0 is very small. Only
eight such objects are currently known
(Tab.\,\ref{tab:hotdas}\footnote{We omitted from this list WD0316+002,
  which had formally been assigned
  $T\mathrm{\hspace*{-0.4ex}_{eff}}$/\logg\ = 140\,000/8.0 by
  \citet{2011ApJ...730..128T}, because the determined temperature was
  outside of their model atmosphere grid. Other investigations gave
  \Teff\ well below 100\,000\,K \citep[see, e.g.,
  ][]{2013ApJS..204....5K}. We also omitted PG1342$+$444, for which
  \citet{2011ApJ...730..128T} found 104\,840/7.71, but
  \citet{2011ApJ...743..138G} and \citet{2014MNRAS.440.1607B} claim
  \Teff\ around 70\,000\,K. One new candidate with 151\,505/7.61
  \citep[SDSS\,J145545.58+041508.6; ][]{2019MNRAS.482.5222T} emerged
  from the most recent analysis of DAs in the Sloan Digital Sky Survey (data
  release 14) and deserves further studies.}). This paucity of
extremely hot H-rich WDs is in stark contrast to the much larger
number of forty hydrogen-poor (non-DA) WDs with \Teff $\geq$
100\,000\,K and \logg $\geq$ 7, that are 14 helium-rich (DO)
WDs\footnote{See compilation by N. Reindl:
  \url{https://www.star.le.ac.uk/~nr152/} and Reindl et al., in prep.}
and 26 He- and C-rich (PG1159) stars (see \citet{2006PASP..118..183W}
and Werner et al., in prep.). Hence, the DA/non-DA ratio at the
hottest end of the WD cooling sequence is 0.2, while for the cooler
WDs, that ratio is near five \citep{2011ApJ...737...28B}. In
Fig.\,\ref{fig:gteff} we depict the location of all known WDs with
\Teff $\geq$ 100\,000\,K and \logg $\geq$ 7.0 in the $g$ --
\Teff\ diagram.

\begin{table*}
\begin{center}
\caption{Currently known hydrogen-rich WDs with \Teff $\geq$
  100\,000\,K and \logg $\geq$ 7.0. All results were obtained with NLTE
  models. For the three hottest WDs, Balmer line fits were performed,
  while for the others metal lines were exploited to constrain
  \Teff. For the star analyzed in the present paper, \pg, we quote the
  latest literature result (for earlier results and the values derived here
  see Table~\ref{tab:earlierresults}).}
\label{tab:hotdas} 
\small
\begin{tabular}{llllllll}
\hline 
\hline 
\noalign{\smallskip}
Name & WD name &  Type & \Teff/K & \logg  & Technique   & Reference \\ 
\hline 
\noalign{\smallskip}
EGB 1      & WD0103$+$732 & DAO & 147\,000 $\pm$ 25\,000 & 7.34 $\pm$ 0.31 & NLTE Balmer & {\citet{1999A&A...350..101N}}\\
WeDe 1     & WD0556$+$106 & DA  & 141\,000 $\pm$ 19\,000 & 7.53 $\pm$ 0.32 & NLTE Balmer & {\citet{1999A&A...350..101N}}\\
\pg        & WD0948$+$534 & DA  & 141\,000 $\pm$ 12\,000 & 7.46 $\pm$ 0.18 & NLTE Balmer & {\citet{2017ASPC..509..195P}}; see Table~\ref{tab:earlierresults}\\
Longmore 1 & --           & DAO & 120\,000 $\pm$ 10\,000 & 7.00 $\pm$ 0.30 & NLTE metals & {\citet{2012zieglerdiss}} \\
Abell 31   & WD0851$+$090 & DAO & 114\,000 $\pm$ 10\,000 & 7.40 $\pm$ 0.30 & NLTE metals & {\citet{2012zieglerdiss}} \\
Abell 7    & WD0500$-$156 & DAO & 109\,000 $\pm$ 10\,000 & 7.00 $\pm$ 0.30 & NLTE metals & {\citet{2012zieglerdiss}} \\
EGB 6      & WD0950$+$139 & DAO & 105\,000 $\pm$ \hspace{0.72mm} 5\,000 & 7.40 $\pm$ 0.40 & NLTE metals & {\citet{2018A&A...616A..73W}} \\
NGC 3587   & WD1111$+$552 & DAO & 104\,000 $\pm$ 10\,000 & 7.00 $\pm$ 0.60 & NLTE metals & {\citet{2012zieglerdiss}} \\
\noalign{\smallskip} \hline
\end{tabular} 
\end{center}
\end{table*}

This phenomenon has been discussed in the literature since decades,
when a lack of very hot DAs in the Palomar-Green (PG) survey
\citep{1986ApJS...61..305G} was reported \citep{1986ASSL..128..367L}.
For example, the central star of the planetary nebula WeDe~1 (= WDHS~1)
was initially regarded as the only hydrogen-rich WD with a
``PG1159-like'' \Teff\ and gravity (about 140\,000\,K and \logg = 7.0) by
\cite{1994ApJ...424..817L}. The authors emphasized, however, that the
parameters of the H-rich WDs remain uncertain because of unknown
element abundances and the so-called Balmer-line problem. It was
speculated that the improvement of model atmospheres might show that
the number of DAs and DAOs increases substantially. The basic problem
is the large uncertainty when the analyses solely rely on optical
spectra. In the case of DAs, only the Balmer lines are available and
in addition \ion{He}{ii} $\lambda$4686 in the DAOs. A further
complication is the appearance of the Balmer-line problem for many of
the hottest DAs \citep{1992LNP...401..310N}, meaning that for a
particular object different temperatures follow from fits to different
Balmer line series members. This problem is still not fully
understood. To some extent it is due to the neglect of metal opacities
in the models \citep{1996ApJ...457L..39W}, but it was found in many
cases that the problem does not disappear even if sophisticated
metal-line blanketed models are used \citep[e.g.,
][]{2018A&A...616A..73W}.

While initially LTE model
atmospheres were used for spectral analyses, the subsequent employment
of non-LTE models was more appropriate considering the high
temperatures. But the lack of a more precise temperature indicator,
the ionization balance of any species, remained and left analyses
based on Balmer lines with high uncertainties. In the case of
non-DAs, lines from highly ionized C, N, O, and Ne are seen in the
optical spectra of many objects, so the situation is better for these
stars and analysis results are more reliable. The only possibility to
improve the precision of parameter estimations of the hottest DA(O)s
is ultraviolet (UV) spectroscopy. Here, lines from many species in
different ionization stages are usually detected. 

Analyses of hot DA(O)s employing UV spectroscopy showed that a number
of the hottest objects are in fact even hotter than previously
thought. For example, the temperatures of some of the WDs analyzed by
\citet{1999A&A...350..101N} using Balmer lines, were shown to be above
100\,000\,K when UV spectroscopy was employed. One example is the
central star of Abell~31, for which \Teff = 84\,700\,K was derived from
the Balmer lines, while \citet{2012A&A...548A.109Z} found 114\,000\,K
from a detailed analysis of metals lines in the UV.  In this sense,
the temperatures of the objects listed in Table~\ref{tab:hotdas} that
were determined from Balmer line spectroscopy alone must be regarded
uncertain.  Interestingly, this is the case of the three hottest ones,
which have temperatures around 140\,000\,K. These are the just
mentioned central star of WeDe~1, the central star of EGB~1, and
\pg. From these three stars, UV spectroscopy is only available for
\pg\, therefore we are assessing it in the present study.

\begin{table*}
\begin{center}
\caption{Previous determinations of effective
  temperature and gravity for \pg\ as derived from Lyman and Balmer
  line fits and result of this work. The two different sets of values from
  \citet{2017ASPC..509..195P} using Lyman lines were obtained
  employing different line-broadening tables. The metal abundances
  used in the NLTE model structures computed by
  \citet{2003MNRAS.341..870B}, \citet{2010ApJ...720..581G} and
  \citet{2017ASPC..509..195P} were chosen by an approximative estimate
  (see text and in Table~\ref{tab:abu} lines ``B03 model'', ``G10
  model'', ``P17 model'', respectively). The entry ``(+metals)'' set
  in brackets means that metals were included in the model atmosphere,
  but metal lines were not used to constrain \Teff\ and \logg.}
\label{tab:earlierresults} 
\small
\begin{tabular}{llll}
\hline 
\hline 
\noalign{\smallskip}
 \Teff/K & \logg  & Technique   & Reference \\ 
\hline 
\noalign{\smallskip}
126\,300              & 7.27            & LTE, Balmer              & \citet{1995LNP...443...12L}\\
110\,000 $\pm$ 2\,500   & 7.58 $\pm$ 0.06 & NLTE (+metals), Balmer   & \citet{2003MNRAS.341..870B}\\
130\,370 $\pm$ 6\,603   & 7.26 $\pm$ 0.12 & NLTE, Balmer             & \citet{2005ApJS..156...47L}\\
139\,500              & 7.56            & NLTE (+metals), Balmer   & \citet{2010ApJ...720..581G,2011ApJ...743..138G}\\
138\,000 $\pm$ 7\,000   & 8.43 $\pm$ 0.10 & NLTE (+metals), Lyman    & \citet{2017ASPC..509..195P}; line broadening of {\citet{1997A&AS..122..285L}}\\
140\,000 $\pm$ 8\,000   & 8.60 $\pm$ 0.10 & NLTE (+metals), Lyman    & \citet{2017ASPC..509..195P}; line broadening of {\citet{2009ApJ...696.1755T}}\\
141\,000 $\pm$ 12\,000 & 7.46 $\pm$ 0.18 & NLTE (+metals), Balmer   & \citet{2017ASPC..509..195P}\\
105\,000 $\pm$ 5\,000  & 7.70 $\pm$ 0.20 & NLTE+metals, metal lines, Lyman & This work\\
\noalign{\smallskip} \hline
\end{tabular} 
\end{center}
\end{table*}

\begin{table*}
\begin{center}
\caption{Assumed and derived metal abundances in previous studies
  compared to the result of the present work (last line). Abundances
  given as number ratio relative to hydrogen. B03:
  \citet{2003MNRAS.341..870B}; ``model'' denotes the abundances
  assumed for the model calculation and ``synspec'' denotes abundances
  derived by fitting of metal lines without considering back-reaction
  on population numbers and atmospheric structure (no fits to lines
  from C, N, O, and Si could be achieved). P17:
  \citet{2017ASPC..509..195P}; same procedure, but considering more
  species. G10: \citet{2010ApJ...720..581G}; assumed abundances.}
\label{tab:abu} 
\small
\begin{tabular}{lcccccccccc}
\hline 
\hline 
\noalign{\smallskip}
                     & C               & N               & O               & Si              & P               & S               & Ar               & Fe              &  Ni \\
\hline 
\noalign{\smallskip}
B03 model            &$4.0\cdot 10^{-7}$&$1.6\cdot 10^{-7}$&$9.6\cdot 10^{-7}$&$3.0\cdot 10^{-7}$&              $-$&              $-$&              $-$&$1.0\cdot 10^{-5}$&$5.0\cdot 10^{-7}$\\ 
\hspace{5mm} synspec &              $-$&              $-$&              $-$&              $-$&              $-$&              $-$&              $-$&$1.9\cdot 10^{-6}$&$1.2\cdot 10^{-7}$\\ 
P17 model            &$1.0\cdot 10^{-6}$&$1.0\cdot 10^{-6}$&$1.0\cdot 10^{-5}$&$1.0\cdot 10^{-6}$&$1.0\cdot 10^{-7}$&$1.0\cdot 10^{-7}$&$3.0\cdot 10^{-5}$&$1.0\cdot 10^{-6}$&$1.0\cdot 10^{-6}$\\ 
\hspace{5mm} synspec &$2.8\cdot 10^{-6}$&$9.1\cdot 10^{-7}$&$1.5\cdot 10^{-5}$&$1.5\cdot 10^{-5}$&$9.2\cdot 10^{-8}$&$1.4\cdot 10^{-6}$&$1.1\cdot 10^{-6}$&$1.2\cdot 10^{-5}$&$3.8\cdot 10^{-6}$\\ 

G10 model            &$2.5\cdot 10^{-4}$&$6.0\cdot 10^{-5}$&$4.6\cdot 10^{-4}$&              $-$&              $-$&              $-$&              $-$&$-$&$-$\\  

\noalign{\smallskip}
This work            &$<2.5\cdot 10^{-6}$&$2.2\cdot 10^{-7}$&$3.2\cdot 10^{-8}$&$7.2\cdot 10^{-6}$&$1.6\cdot 10^{-7}$&$1.6\cdot 10^{-6}$&$9.1\cdot 10^{-7}$&$5.4\cdot 10^{-5}$&$6.9\cdot 10^{-6}$\\ 
\noalign{\smallskip} \hline
\end{tabular} 
\end{center}
\end{table*}

With the advent of the Far Ultraviolet Spectroscopic Explorer (FUSE),
it became possible to use the Lyman line series beyond Ly$\alpha$ as
diagnostics for a large number of DA(O)s. However, it turned out that
particularly for the hottest stars, systematic differences between the
results derived from Lyman and Balmer line analyses appeared
\citep{2001MNRAS.328..211B}. This discrepancy is still not understood
\citep{2015ASPC..493...15P}. One example is the DA of the present
study for which parameter estimates widely differ as we will describe
later.

We report here on our analysis of UV spectra of \pg\ taken with FUSE
and the Hubble Space Telescope (HST) in order to find out whether this
WD is indeed one of the three hottest DAs known or if it is
significantly cooler. In the latter case, the paucity of hot
hydrogen-rich WDs relative to the non-DAs would become even more
evident.

We begin with an overview of previous work on this WD in the following
Section. Then we describe our spectral analysis in
Sect.\,\ref{sect:previous}. In Sect.\,\ref{sect:variability} we report
on the variability of the star and we conclude with a summary and
discussion of our results in Sect.\,\ref{sect:results}.

\section{Previous studies of \pg}
\label{sect:previous}

Numerous attempts were made over more than the past two decades to
determine effective temperature and gravity of
\pg\ (Table~\ref{tab:earlierresults}). To understand the significant
scatter in the derived results, it is necessary but also instructive
to report in some detail the used analysis techniques and atmosphere
models. Initially, the star was identified as an extremely hot DA
(\Teff = 126\,000\,K) by \citet{1995LNP...443...12L}, fitting the
Balmer lines with pure-hydrogen LTE models. In a systematic analysis
of WDs from the PG survey \citep{1986ApJS...61..305G} it was found
that \pg\ is by far the hottest DA in that sample \citep[\Teff =
  130\,370\,K,][]{2005ApJS..156...47L}, and the only DA together with
PG\,0950+139, the central star of the planetary nebula EGB\,6
\citep[\Teff =
  105\,000\,K,][]{2005ApJS..156...47L,2018A&A...616A..73W}, that
exceed \Teff = 100\,000\,K. It was also emphasized that this was the
first DA exhibiting the Balmer line problem, a phenomenon up to then
restricted to DAO WDs.

The first spectral analysis using NLTE models was performed by
\citet{2003MNRAS.341..870B}. To account for the effects of
metal opacities on the atmospheric structure and the Balmer lines
\citep{1996ApJ...457L..39W}, they included, as a first approximation,
several species in the calculations (C, N, O, Si, Fe, Ni), assuming
abundances equal to that derived for another hot DA (G191-B2B), see
Table\,\ref{tab:abu}; line ``B03 model''. As a result, the Balmer line
fit arrived at a lower temperature of 110\,000\,K (and \logg =
7.58). Keeping fixed these atmospheric values, Fe and Ni abundances
were derived from UV spectra (Table\,\ref{tab:abu}; line ``B03
synspec'') by varying the original abundance estimates in the formal
solution for the radiation transfer, only (i.e., disregarding
back-reaction on population numbers and atmospheric structure), but
that procedure failed to achieve a fit to lines of C, N, O, and Si in
the UV spectra. A later attempt with chemically stratified models
failed, too \citep{2012MNRAS.421.3222D}.

\begin{figure*}
 \includegraphics[height=3.4cm]{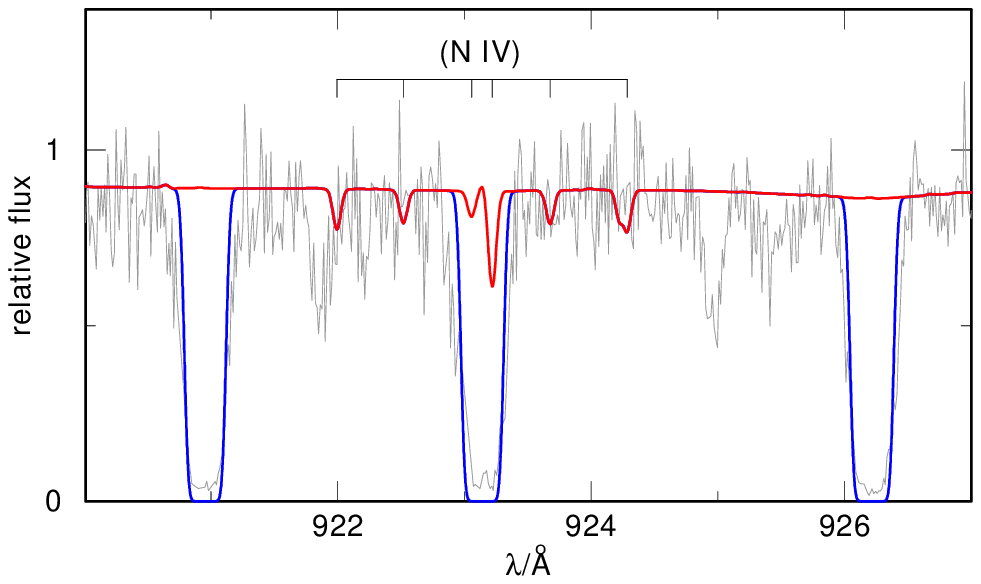}
\hspace{.1cm}
 \includegraphics[height=3.4cm]{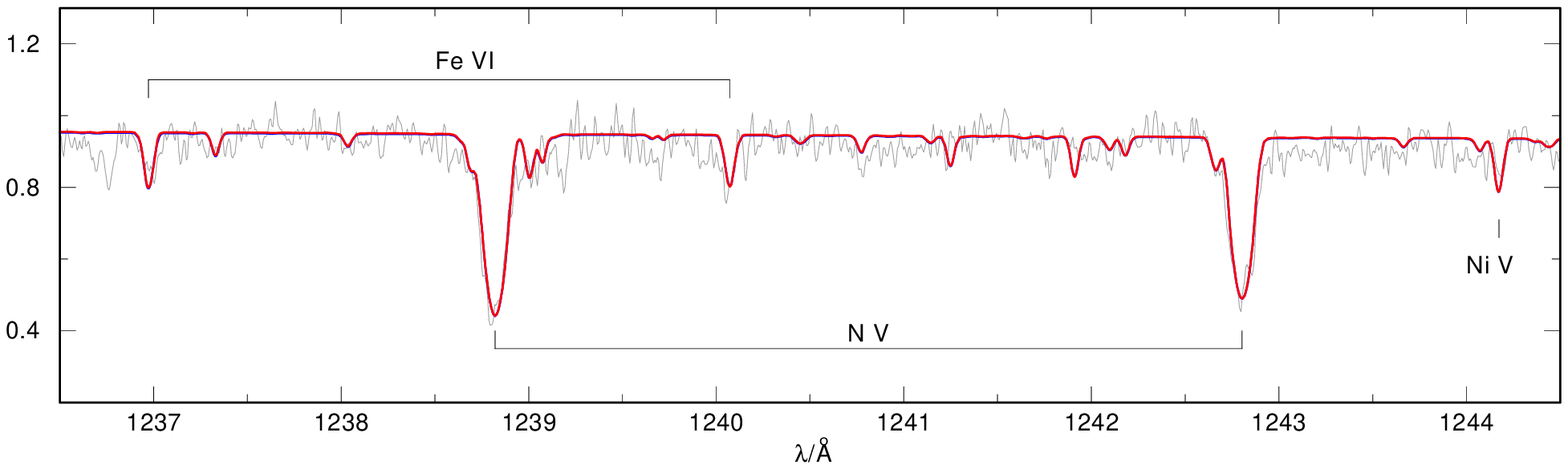}
  \caption{Spectral regions with a \ion{N}{iv} multiplet (left) and the
    \ion{N}{v} resonance doublet (right). \ion{N}{iv} lines in the
    model (red) are not detectable in the observation (grey), thus,
    \Teff $> 100\,000$\,K. The blue graph is our model attenuated by
    simulated ISM Lyman lines.}
\label{fig:niv}
\label{fig:nv}
\end{figure*}

\begin{figure*}
 \includegraphics[height=3.4cm]{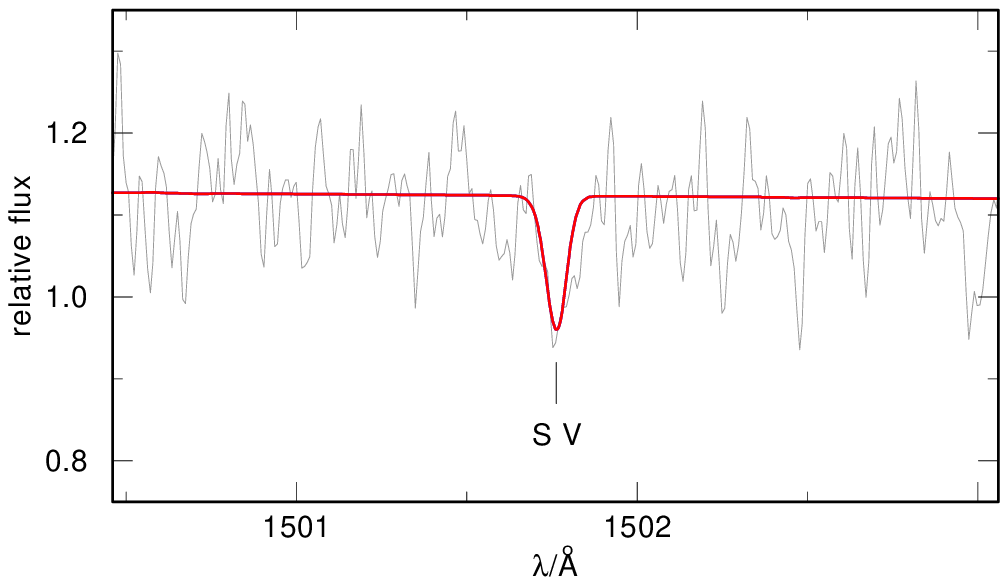}
\hspace{.1cm}
 \includegraphics[height=3.4cm]{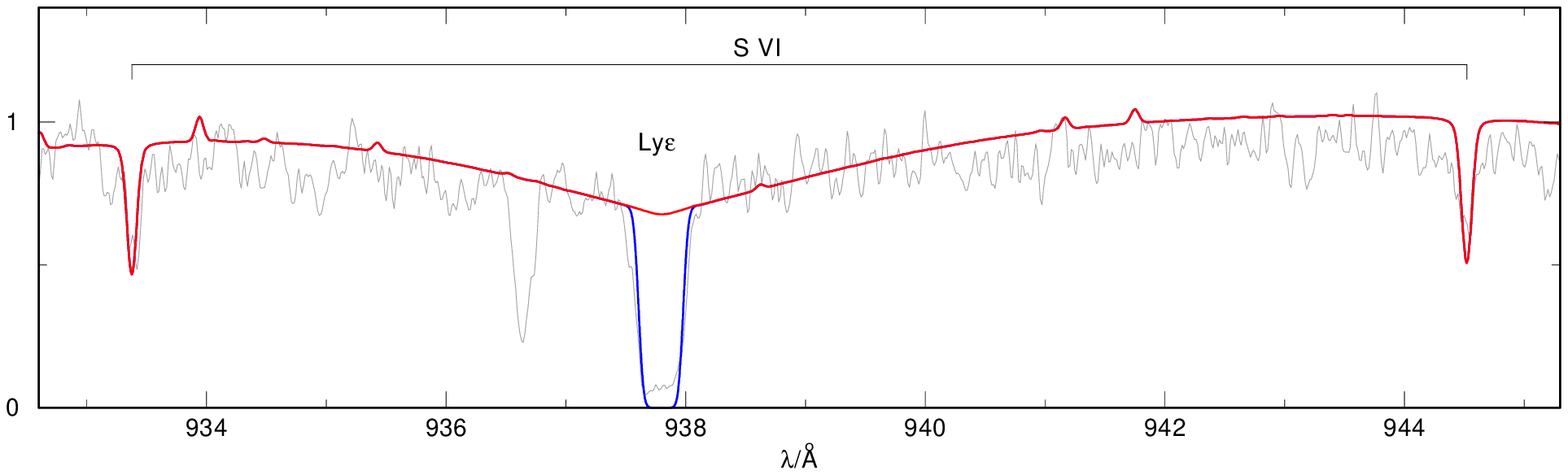}
  \caption{Fits to \ion{S}{v} $\lambda$1502 (left) and the \ion{S}{vi}
    resonance doublet (right).}
\label{fig:sv}
\label{fig:svi}
\end{figure*}

In a subsequent NLTE analysis \citet{2010ApJ...720..581G} derived a
significantly higher temperature of almost 140\,000\,K (and \logg =
7.56) from Balmer lines. The only difference in their model grid
compared to the previous NLTE study was the assumption of other metal
abundances (solar values, including the species C, N, and O; see
Table\,\ref{tab:abu}; line ``G10 model''). The most recent assessment
of the Balmer-line profiles of \pg\ was presented by
\citet{2017ASPC..509..195P}. Their analysis procedure, using a model
with 110\,000\,K, \logg = 7.58, was the same like in
\citet{2003MNRAS.341..870B}, but additional species were included (P,
S, Ar). The abundances derived in an approximate way (again, assuming
an initial estimate, line ``B17 model'' in Table\,\ref{tab:abu}, and
just varying abundances in the formal solutions to find the values
given in line ``B17 synspec'' in Table\,\ref{tab:abu}), were then used
to compute a new model grid to find improved values for \Teff\ and
\logg\ from the Balmer and Lyman lines. From the Balmer lines they
found a much higher temperature (141\,000\,K and \logg = 7.46)
compared to their previous study (110\,000\,K). Most puzzling however,
was the result of their fit to the Lyman lines in the FUSE spectrum
with this model grid. While the same high effective temperature was
derived like for the Balmer lines ($\approx$ 140\,000\,K), an
extremely large value for the surface gravity was determined (\logg =
8.43 -- 8.60, depending in detail on the used line-broadening
tables). This is in strong contrast to the gravity estimates from
their and all previous Balmer line fits, which gave values of
\logg\ in the narrow range 7.26 -- 7.56.

To summarize the findings of the previous studies, rather conflicting
results for \Teff\ (110\,000 -- 141\,000\,K) and \logg\ (7.26 -- 8.60)
were derived. For the present paper, we set out to determine these
parameters more precisely by employing UV metal lines as sensitive
temperature indicators and to determine metal abundances in order to
compute a self-consistent atmosphere structure.

\section{Spectral analysis}
\label{sect:analysis}

\subsection{UV observations}

We used archival spectra of \pg\ taken with FUSE and HST. Three
observations were performed with FUSE (Data IDs A0341501000, --2000,
and --3000) in 2007 and 2008 with a total exposure time of
16\,218\,s. However, in the first and second of these observations, data
of two channels are missing while in the third observation all four
channels are complete. We co-added the available spectra. The useful
range is 910--1180\,\AA. The star was also observed with the Space
Telescope Imaging Spectrograph (STIS) aboard HST during four
consecutive orbits on April 20, 2000, using grating E140M at
central wavelength 1425\,\AA\ (Datasets O59P05010, --20, --30,
--40). The co-added spectrum has a total exposure time of 11\,958\,s and
was retrieved from the
StarCAT\footnote{\url{https://archive.stsci.edu/prepds/starcat/}}
catalogue \citep{2010ApJS..187..149A}. Its useful wavelength range is
1150--1710\,\AA. 

Considering the optical variability of the star with an amplitude of
$\Delta V = 0.19$\,mag and a period of 3.45\,d
(Sect.\,\ref{sect:variability}), we inspected the UV flux for
variability. The four sub-exposures of the STIS observation were taken
within a time span of just 6\% of the variability period.  The flux
levels of the three FUSE exposures do not differ by more than
10\,\%. Also, there is no significant difference in the flux levels of
the co-added FUSE spectrum and the HST spectrum in the wavelength
region where they overlap. A comparison of the HST spectrum with a
low-resolution spectrum taken with the International Ultraviolet
Explorer (image SWP14328) in 1981 also shows no obvious difference. 

In all respective figures displayed in this paper, the photospheric
lines were shifted to rest wavelengths. The photospheric lines in the
HST spectrum are blueshifted by $v_{\rm phot}$ = $-10$\,km/s, as
indicated by, e.g., the \ion{N}{v} resonance doublet, \ion{O}{v}
$\lambda$1371, \ion{S}{v} $\lambda$1502, and several \ion{Fe}{vi}
lines. \citet{2012MNRAS.423.1397D} derived, from the same HST
spectrum, a significantly higher value of $-17$\,km/s. This can be
traced back to the fact that they assigned the strong \ion{C}{iv}
resonance doublet (at this redshift) to the photosphere, but we will
show below that it is in fact dominated by an interstellar absorption
component. The photospheric lines in the FUSE spectrum are blueshifted
by $v_{\rm phot} \approx -25$\,km/s, but we note that the absolute
wavelength calibration of the FUSE data is not as reliable as that of
STIS.  To account for the FUSE spectral resolving power (R = 20\,000),
the model spectra were convolved with a 0.05\,\AA\ (FWHM) Gaussian. In
some cases, the observations were slightly smoothed (with up to
0.03\,\AA\ Gaussians). The model spectra fitted to the STIS
observations were convolved with Gaussians to account for the
resolution (R = 38\,000), i.e., their FWHM is ranging from 0.030 to
0.045\,\AA\ over the spectral window 1150--1710\,\AA.

\begin{figure*}
 \includegraphics[height=5cm]{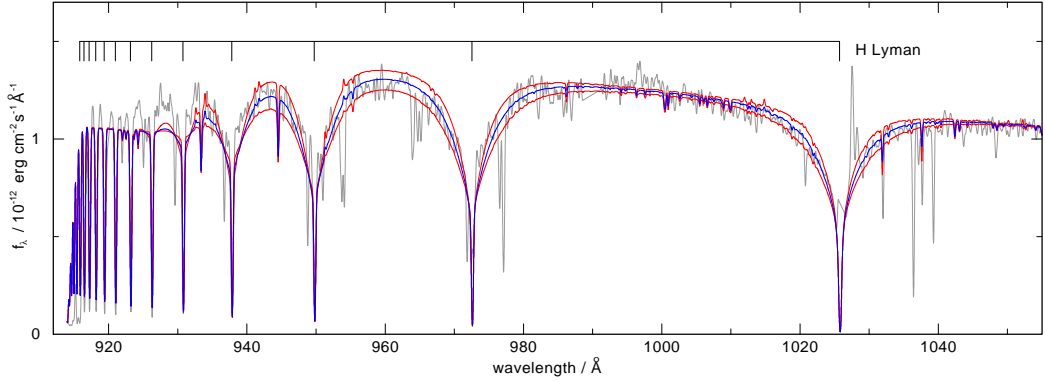}
  \caption{Observed Lyman lines from Ly$\beta$ to the series limit,
    compared to three models with different surface gravity, \logg =
    7.7 (blue graph) and $\pm$0.5~dex (red graphs). \Teff\ and element
    abundances according to final model
    (Table~\ref{tab:results}). Model fluxes are normalized to the
    observed continuum at 1050\,\AA, reddened ($E(B-V) = 0.02$), and
    attenuated by ISM interstellar neutral H absorption ($n_{\rm HI}= 2.4
    \times 10^{19}$\,cm$^{-2}$), the latter affecting the innermost Ly
    line cores, only. For clarity, observation and models are smoothed
    by 0.2\,\AA\ Gaussians.}
\label{fig:logg}
\end{figure*}

\begin{figure}
 \centering\includegraphics[width=0.8\columnwidth]{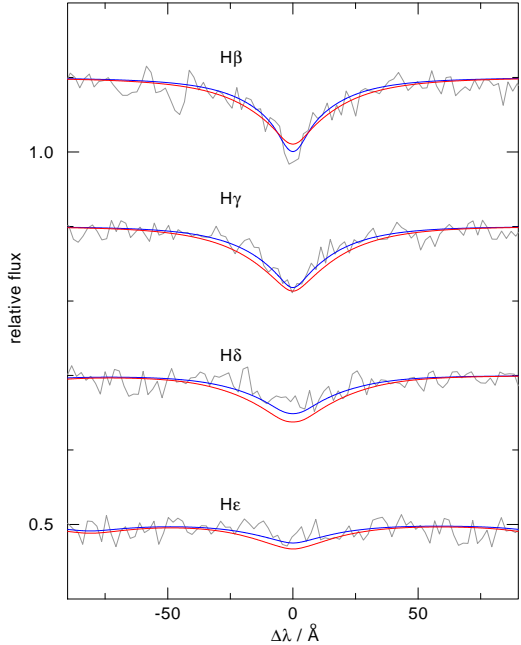}
  \caption{Observed Balmer lines compared to our final model (blue graph) and a
    model with the same parameters but without metals (red).}
\label{fig:balmer}
\end{figure}

\subsection{Model atmospheres}

We used the T\"ubingen Model-Atmosphere Package
(TMAP\footnote{\url{http://astro.uni-tuebingen.de/~TMAP}}) to compute
non-LTE, plane-parallel, line-blanketed atmosphere models in radiative
and hydrostatic equilibrium
\citep{1999JCoAM.109...65W,2003ASPC..288...31W,tmap2012}. The models
include H, He, C, N, O, Si, P, S, Fe, and Ni. The employed model atoms
were described in detail by \citet{2018A&A...609A.107W}. In addition,
we performed line formation iterations (i.e., keeping  the atmospheric
structure fixed) for argon using the model atom presented in
\citet{2015A&A...582A..94W}. Since the treatment of Lyman line
broadening was a topic in previous works, we mention that we used the
broadening tables of \citet{2009ApJ...696.1755T}.

\begin{figure*}
 \includegraphics[height=3.4cm]{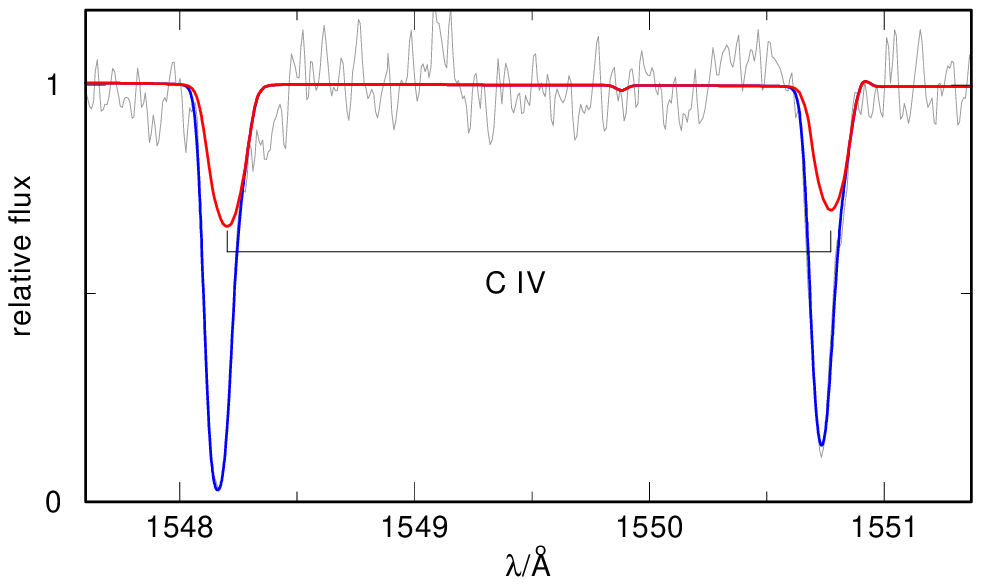}
\hspace{.1cm}
 \includegraphics[height=3.4cm]{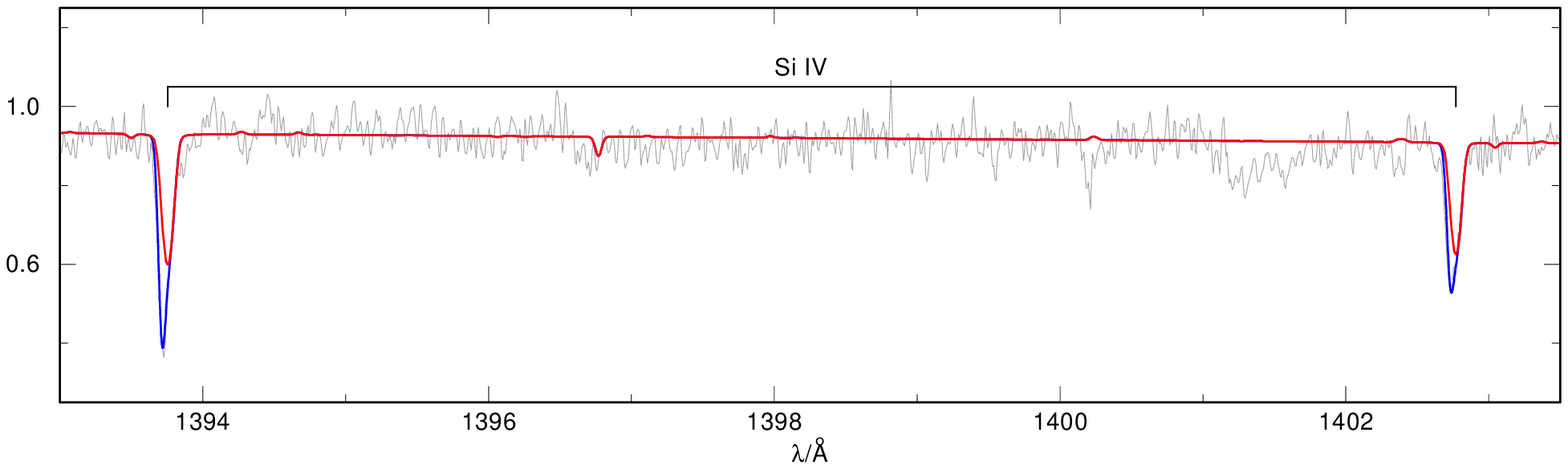}
  \caption{Left panel: The observed
    \ion{C}{iv} resonance doublet is dominated by ISM absorption
    (fitted by the blue graph). The weaker photospheric contribution
    is indicated by the model (red graph) computed with the upper
    abundance limit for C derived from the absence of subordinate
    \ion{C}{iv} lines. Right panel: The computed \ion{Si}{iv} resonance doublet
    has an ISM (blue graph) and a photospheric contribution (red
    graph).}
\label{fig:siiv}
\label{fig:civ}
\end{figure*}

\begin{figure*}
 \includegraphics[height=3.4cm]{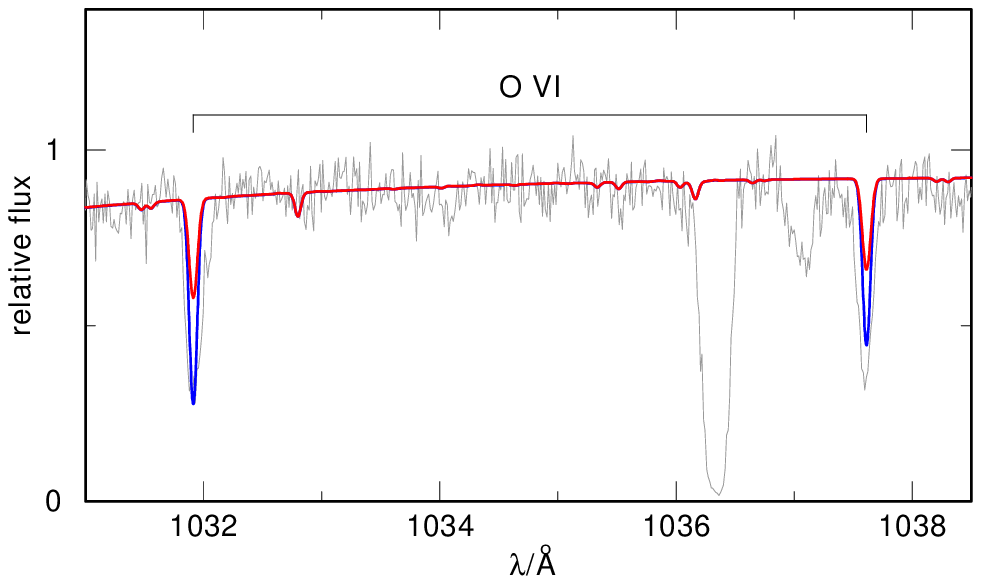}
\hspace{.1cm} \includegraphics[height=3.4cm]{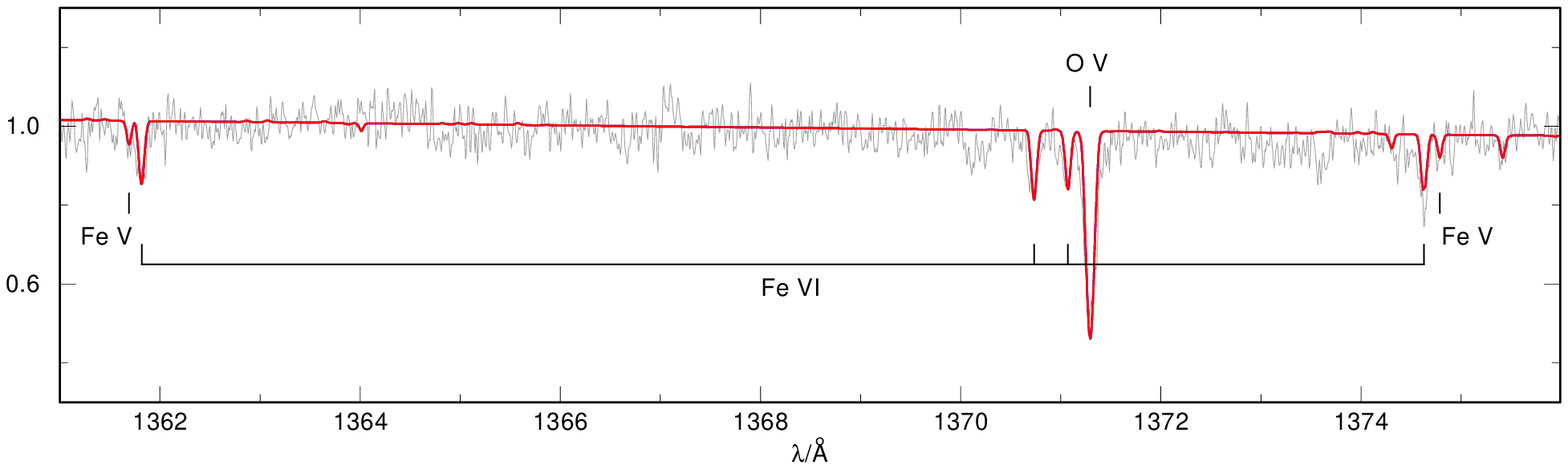}
  \caption{Fits to the \ion{O}{vi} resonance doublet (left) and
    \ion{O}{v} $\lambda$1371 (right). The \ion{O}{vi} doublet is not
    matched by the model and could be dominated by ISM contribution
    (blue graph).}
\label{fig:ov}
\label{fig:ovi}
\end{figure*}

\subsection{Effective temperature and surface gravity}

To find a model that best fits the metal lines in the UV spectra, we
followed a manual, iterative procedure to improve the fits by
computing a sequence of models successively, changing effective
temperature and abundances. At the outset, we kept fixed the surface
gravity to \logg = 7.7, a value guided by the Balmer-line fits in the
literature (Table\,\ref{tab:earlierresults}). To constrain \Teff, the
relative strengths of lines from different ionization stages of a
given chemical element were used. Most useful is the fact, that
\pg\ displays lines from four ionization stages of iron, namely
\ion{Fe}{v-viii}. This immediately constrains the temperature to the
narrow range of 100\,000 -- 110\,000\,K. Cooler and hotter models do not
exhibit the lines detected from \ion{Fe}{viii} and \ion{Fe}{v},
respectively. Ionization balances from other elements are in
accordance with this temperature range. While the \ion{N}{v} resonance
doublet depends weakly on \Teff, the \ion{N}{iv} lines are very
sensitive. The non-detection of the \ion{N}{iv} multiplet near
923\,\AA\ (Fig.\,\ref{fig:nv}) requires \Teff$>$100\,000\,K. On the
other hand, sulfur gives a strong upper limit. At 110\,000\,K the
\ion{S}{v} 1502\,\AA\ line (Fig.\,\ref{fig:sv}) is becoming too weak,
while at 105\,000\,K it fits well, together with the \ion{S}{vi}
resonance doublet. We conclude \Teff = 105\,000$\pm$5000\,K. The only
serious problem is posed by the \ion{O}{vi} resonance doublet, which
is much too weak in the model, when \ion{O}{v} $\lambda$1371 fits
well. This mismatch between the \ion{O}{v} and \ion{O}{vi} lines
cannot be mitigated with models of different temperature within
reasonable limits. We will discuss below, that an ISM contribution to
\ion{O}{vi} could be an explanation for this discrepancy.

Having found the best fitting model in this way, we then considered
deviations from the assumed value of \logg = 7.7. We computed models
with different gravity, keeping fixed all other model parameters and
compared the resulting Lyman lines with the observations, with the
exception of Ly$\alpha$, which is dominated by ISM absorption over the
entire profile. The higher Lyman series members are dominated by ISM
contribution in the innermost cores only.

In Fig.\,\ref{fig:logg} we compare model profiles of Ly$\beta$ and
higher series members with the observation. We show our final model
(\Teff = 105\,000\,K, \logg = 7.7) together with two models with
gravity increased and decreased by 0.5~dex. Obviously, the line
profiles do not depend strongly on gravity, so that we must accept a
relatively large error of $\pm 0.4$ dex. As we shall see
below (Sect.\,\ref{sect:para}), the parallax of \pg\ favors \logg =
7.6, so that we finally adopt a smaller error of $\pm 0.2$ dex. It can
be seen that the continuum flux level at the highest series members at
wavelengths shorter than about 930\,\AA\ is not quite matched by our
models. In fact, it is almost identical in all three models, which
means that level dissolution has essentially eliminated photospheric
line opacities (the narrow line features are interstellar
absorptions). The flux below this wavelength threshold
is therefore unaffected by line opacities and thus determined by the
effective temperature. So a slight increase in \Teff\ to 110\,000\,K
leads to a better continuum flux fit, but this is within our error
range for the temperature. Note that the flux cut-off at 912\,\AA\ is
entirely due to interstellar neutral H absorption, which is included
in the models shown in Fig.\,\ref{fig:logg} ($n_{\rm HI}= 2.4 \times
10^{19}$\,cm$^{-2}$, Sect.\,\ref{sect:reddening}).

We also assessed the question why \citet{2017ASPC..509..195P} could
arrive at extraordinary high gravities of around \logg = 8.6 in their
Lyman line fitting. We computed a model with that gravity and \Teff =
140\,000\,K, i.e., the parameters they derived, and compared it to our
105\,000/7.7 model. It turns out that the effect of an increase in
\Teff\ on the Lyman line profiles can to a large extent be compensated
by an increase in \logg. Both models show differences in the wings of
Ly$\beta$ to Ly$\delta$ that are smaller than a few percent, only. So,
even small errors in the flux calibration can lead to large errors in
the determination of \Teff\ and \logg. We also note that we have
included interstellar reddening ($E(B-V) = 0.02$,
Sect.\,\ref{sect:reddening}) in the model fluxes displayed in
Fig.\,\ref{fig:logg}, which does affect the continuum shape and must
therefore be considered in flux fitting across this wide wavelength
range. We will show below (Sect.\,\ref{sect:para}), that the parallax
of \pg\ independently excludes high gravities.

In the final model calculations, helium is included with an abundance
value identical to the upper limit determined from the absence of
\ion{He}{ii} $\lambda$1640, i.e., He $= 3\times10^{-4}$ (by mass).

We compared an optical spectrum of \pg\ covering H$\beta$ and higher
Balmer series members with our final model and noticed that the
Balmer-line problem is still present; the H$\beta$ line core of the
model is not deep enough (Fig.\,\ref{fig:balmer}). The influence of C, N, and O on the model
structure is relatively weak because of the strongly subsolar
abundances of these species (see next Section).

\begin{figure*}
 \includegraphics[height=3.4cm]{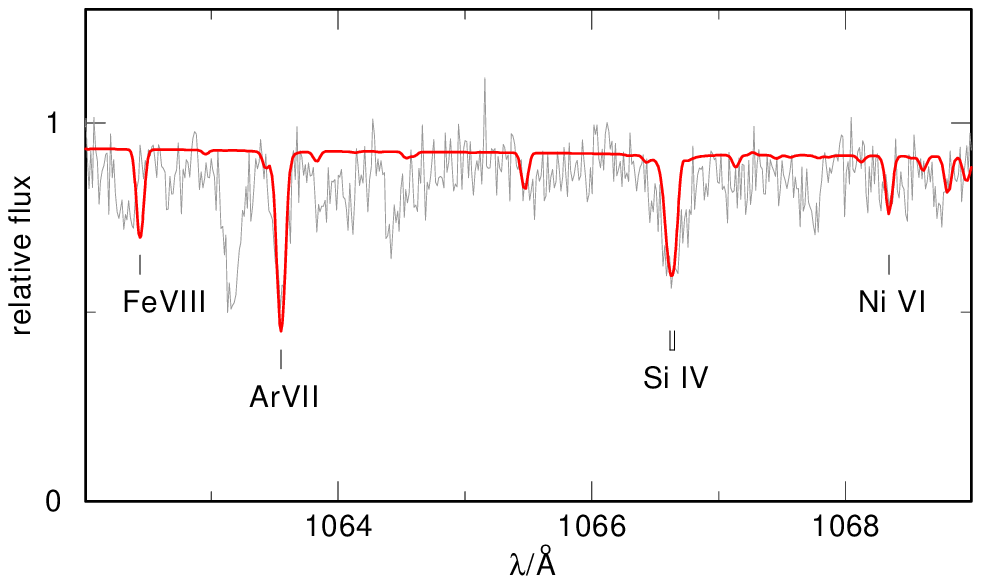}
\hspace{.1cm}
 \includegraphics[height=3.4cm]{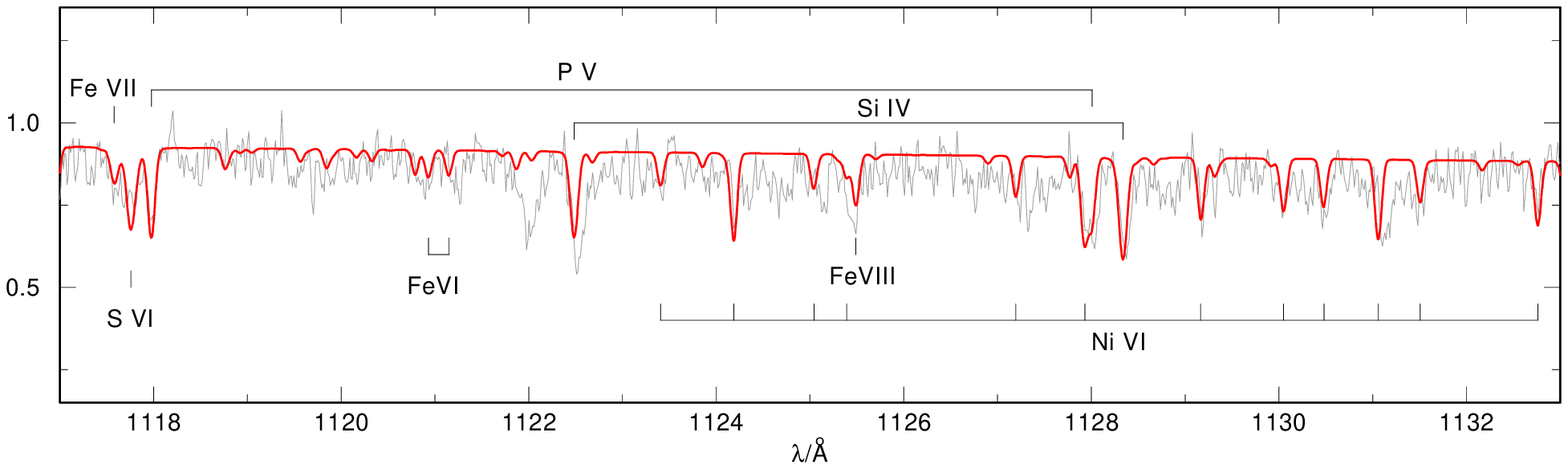}
  \caption{Left: Lines from \ion{Si}{iv}, \ion{Ar}{vii},
    \ion{Fe}{viii}, and \ion{Ni}{vi}. Right: Lines from \ion{Si}{iv},
    \ion{P}{v}, \ion{S}{vi}, \ion{Fe}{vi-viii}, and \ion{Ni}{vi}.}
\label{fig:sips}
\label{fig:arvii}
\end{figure*}

\subsection{Element abundance determination}

Element abundances were derived from line profile fits as indicated in
the following. All quoted abundance values are given in mass fractions.

\subsubsection{Carbon, nitrogen, and oxygen}

Carbon lines cannot be identified. An upper abundance limit of C = $3
\times 10^{-5}$ is obtained from the absence of the \ion{C}{iv}
$\lambda$1168.8/1169.0 3d--4f doublet in the HST spectrum. The FUSE
spectrum at this position is contaminated by a \ion{He}{i}
$\lambda$1168.67 airglow line (second order from
$\lambda$584.33). This C limit is in accordance with the absence of
the \ion{C}{iv} 1107.6/1107.9 3p--4d doublet. The weak absorption
features visible in the FUSE spectrum at these wavelengths are due to
\ion{Ni}{vi} lines. Looking at the observed \ion{C}{iv} resonance
doublet (Fig.\,\ref{fig:civ}) reveals that it must be dominated by
interstellar absorption (blueshifted by 8\,km/s relative to the
photospheric component). The computed photospheric line is much
weaker. The C upper abundance limit is compatible with the observed
line profile. A significantly higher abundance would be detectable in
the red wings of the interstellar line profiles. 

The only detectable nitrogen lines are the \ion{N}{v} resonance doublet
(Fig.\,\ref{fig:nv}). There is no indication for a blueshifted ISM
component as in the case of the \ion{C}{iv} resonance doublet. As
mentioned above, the absence of the \ion{N}{iv} multiplet near
923\,\AA\ (Fig.\,\ref{fig:nv}) imposes a lower limit for \Teff. 

\ion{O}{v} $\lambda$1371 and the \ion{O}{vi} $\lambda\lambda$1032/1038
resonance doublet are very prominent lines (Fig.\,\ref{fig:ov}). The
oxygen abundance is derived by fitting the \ion{O}{v} line because no
interstellar contribution is expected. With this abundance, the
\ion{O}{vi} doublet is much too weak. The observed \ion{O}{v} line is
weaker than the \ion{O}{vi} doublet. In order to reduce the relative
\ion{O}{v}/\ion{O}{vi} line strength in the model to the observed extent,
\Teff\ would need to be as high as 140\,000\,K. Another option would
be a significantly lower \Teff\ of about 90\,000\,K, when the
temperature becomes too low to keep the \ion{O}{v} line strong,
whereas the \ion{O}{vi} resonance doublet remains more prominent. But
these temperature extremes are excluded by ionization balances of
other metals (N, S, Fe, Ni), see above. 

A possible explanation for this discrepancy could be a dominant ISM
contribution to the \ion{O}{vi} doublet, similar to the case of the
\ion{C}{iv} doublet. The FUSE spectral resolution is not sufficient to
detect such a component with relative blueshift of
8\,km/s. Assuming an interstellar column density of $n_{\rm OVI}= 5.0
\times 10^{13}$\,cm$^{-2}$ gives a good fit to the observed line
strengths. Similar $n_{\rm OVI}$ values were found towards a number of
WDs in the sample investigated by \citet{2010ApJ...723.1762B} to probe
the \ion{O}{vi} content in the local ISM.

\subsubsection{Silicon, phosphorus, sulfur, and argon}

The silicon abundance was found from two \ion{Si}{iv} doublets, namely
the 3p--3d lines at 1122.5/1128.3\,\AA\ (Fig.\,\ref{fig:sips}), and
the 3s--3p resonance doublet at
1393.8/1402.8\,\AA\ (Fig.\,\ref{fig:siiv}). Similar to \ion{C}{iv},
the photospheric \ion{Si}{iv} resonance doublet is blended by an
interstellar component with a relative blueshift of 9\,km/s. Another
\ion{Si}{iv} line, the 3d--4f transition at 1066.6\,\AA, is detected
(Fig.\,\ref{fig:arvii}).

The phosphorus abundance was found from fitting the \ion{P}{v}
$\lambda\lambda$1118/1128 resonance doublet (Fig.\,\ref{fig:sips}).

A singlet line from \ion{S}{v} at 1502\,\AA\ (Fig.\,\ref{fig:sv}, left
panel) and the \ion{S}{vi} $\lambda\lambda$933/945 3s--3p resonance
doublet (Fig.\,\ref{fig:svi}, right panel) were used to infer the
sulfur abundance. Two weaker \ion{S}{vi} lines are seen at
1000.5\,\AA\ (4d--5f) and at 1117.8\,\AA\ (4f--7g,
Fig.\,\ref{fig:sips}).

We have previously announced the detection of the strong \ion{Ar}{vii}
$\lambda$1063.55 line, which is the counterpart of the above-mentioned
$\lambda$1502 line in the isoelectronic \ion{S}{v}
\citep{2007A&A...466..317W}. The argon abundance we derived here (Ar =
$3.6\times 10^{-5}$) is in agreement with our earlier work ($7.9\times
10^{-5}$) within error limits, though not identical because our
present model atmosphere is slightly cooler, and includes more species
and has more realistic metal abundances.

\subsubsection{Iron and nickel}

We detected a large number of lines from \ion{Fe}{vi}, some from
\ion{Fe}{vii}, and a few from \ion{Fe}{viii} and \ion{Fe}{v}, see examples in
Figs.\,\ref{fig:nv}, \ref{fig:ov}, and \ref{fig:arvii}. The iron
ionization balance is very temperature sensitive (see above) and we
obtained a good fit to the lines of all four observed ionization stages
at a slightly oversolar iron abundance. 

We identified a large number of nickel lines from ionization
stages \ion{Ni}{v} (Fig.\,\ref{fig:nv}) and \ion{Ni}{vi}
(Fig.\,\ref{fig:sips}). Like for iron, we derived a slightly oversolar
nickel abundance.

\begin{table}
\begin{center}
\caption{Results of spectroscopic analysis of \pg\ and derived
  parameters. Abundances in number ratios relative to
  H (column 2), in mass fractions (column 3), and logarithmic mass
  fractions relative to solar value \citep[column 4; solar abundances
    from][]{2009ARA&A..47..481A}. Error limits for metal abundances
  are $\pm 0.5$\,dex.}
\label{tab:results} 
\begin{tabular}{crrr}
\hline 
\hline 
\noalign{\smallskip}
\Teff/\,K & $105\,000 \pm 5000$ \\
\noalign{\smallskip}
$\log$($g$/cm/s$^2$) & $7.7 \pm 0.2$      \\
\noalign{\smallskip}
$E(B-V)$  & $0.02\pm0.01$\\
\noalign{\smallskip}
$n_{\rm H}$/cm$^{-2}$ & $(2.4\pm0.2) \times 10^{19}$\\
\noalign{\smallskip}
$M\,/\,M_{\odot}$ & $0.66^{+0.06}_{-0.06}$ \\
\noalign{\smallskip}
$d$\,/\,pc      & $361^{+93}_{-74}$ \\
\noalign{\smallskip}
\hline 
\noalign{\smallskip}
abundances & $N_i/N_H$            & $X_i$               & [$X_i$]\\
\hline 
\noalign{\smallskip}
H         & 1                    & 0.996               &  0.13 \\
He        & $<7.6 \times 10^{-5}$ &$<3.0 \times 10^{-4}$ &$<-2.92$ \\
C         & $<2.5 \times 10^{-6}$ &$<3.0 \times 10^{-5}$ &$<-1.86$ \\ 
N         & $ 2.2 \times 10^{-7}$ & $3.0 \times 10^{-6}$ & $-2.31$ \\ 
O         & $ 3.2 \times 10^{-8}$ & $5.0 \times 10^{-7}$ & $-4.03$ \\ 
Si        & $ 7.2 \times 10^{-6}$ & $2.0 \times 10^{-4}$ & $-0.52$ \\ 
P         & $ 1.6 \times 10^{-7}$ & $5.0 \times 10^{-6}$ & $-0.02$ \\ 
S         & $ 1.6 \times 10^{-6}$ & $5.0 \times 10^{-5}$ & $-0.81$ \\ 
Ar        & $ 9.1 \times 10^{-7}$ & $3.6 \times 10^{-5}$ & $-0.31$ \\ 
Fe        & $ 5.4 \times 10^{-5}$ & $3.0 \times 10^{-3}$ &  0.42 \\ 
Ni        & $ 6.9 \times 10^{-6}$ & $4.0 \times 10^{-4}$ &  0.74 \\ 
\noalign{\smallskip} \hline
\end{tabular} 
\end{center}
\end{table}

\begin{figure}
 \centering\includegraphics[width=1.0\columnwidth]{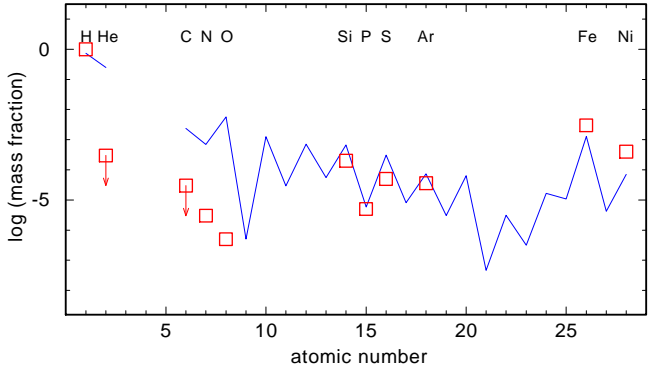}
  \caption{Measured element abundances in \pg\ (red squares) in comparison to
solar abundances (blue line).}
\label{fig:abu}
\end{figure}

\subsubsection{Reddening and ISM neutral hydrogen column density}
\label{sect:reddening}

From a comparison of the overall UV flux distribution of \pg\ with our
final model we found a reddening of $E(B-V) = 0.02\pm0.01$, in
agreement with the 3D dust map of \citet{2018MNRAS.478..651G}, which
gives $E(B-V) = 0.02\pm0.02$ at the Gaia distance
(Sect.\,\ref{sect:para}) of the star. The Ly$\alpha$ line profile is
dominated by ISM absorption. We derived an interstellar neutral
hydrogen column density of $n_{\rm HI}= (2.4 \pm 0.2) \times
10^{19}$\,cm$^{-2}$.

\subsection{Summary of spectral analysis}

The results of our analysis are summarized in Table~\ref{tab:results}
and in Fig.\,\ref{fig:abu}.  We determined \Teff = 105\,000 $\pm$
5000\,K and \logg = 7.7 $\pm$ 0.2. The temperature is lower than that
of all previous analyses, which were in the range 110\,000 --
141\,000\,K. The gravity is in accordance with previous studies (7.27
-- 7.58) when the extreme values of 8.43 -- 8.60 found from Lyman line
fits \citep{2017ASPC..509..195P} are excluded.

The abundances of He, C, N, and O are strongly subsolar by about 2 --
4 dex (upper limits for C and He). Slight underabundances of Si, P, S,
and Ar were found (0.02 -- 0.8 dex) and slight overabundances of Fe
and Ni (0.42 and 0.74 dex). The main difference to the earlier work by
\citet{2017ASPC..509..195P} is our much lower value of the oxygen
abundance (reduced by 2.7 dex), because we concluded that the
\ion{O}{vi} resonance doublet is probably dominated by an ISM
contribution. We also found that the carbon abundance determined by
\citet{2017ASPC..509..195P} must be regarded as an upper limit,
because we showed that the \ion{C}{iv} resonance doublet is dominated
by an ISM contribution.

\subsection{WD mass and distance}
\label{sect:para}

The stellar mass is estimated from the comparison of the determined
atmospheric parameters \Teff\ and \logg\ with evolutionary tracks by
\citet{2016A&A...588A..25M}. We found
$M\,/\,M_\odot = 0.66^{+0.06}_{-0.06}$. The errors are dominated by the
uncertainty in $g$.

The spectroscopic distance $d$ was found by comparing the dereddened
visual magnitude $V_0$ with the respective model atmosphere flux,
resulting in the relation
$$ d {\rm [pc]}= 7.11\times 10^{4} \sqrt{H_\nu\cdot M\cdot 10^{0.4
    V_0-\log g}}\ ,$$
where $H_\nu= 1.54\times 10^{-3}$\,erg cm$^{-2}$s$^{-1}$Hz$^{-1}$ is
the Eddington flux of the model at 5400\,\AA, and $M$ is the stellar
mass in $M_\odot$. For our WD we have $V = 15.32$ mag
\citep{2009ApJ...696..870D}, but note that this is an average value
since it is variable by about $\pm 0.1$ mag (see
Sect.\,\ref{sect:variability}). From $E(B-V)=0.02$ we derived the
visual extinction using the standard relation $A_V=3.1 \times E(B-V) =
0.062$, hence $V_0=15.26$. We
found $d=361^{+93}_{-74}$\,pc and the main error source is our
uncertainty in the gravity (\logg = $7.7\pm0.2$). The distance to the
WD given in the Gaia Data Release 2 ($318\pm7$\,pc) is in good
agreement. A perfect match would require \logg = 7.6, while on the
other hand the unrealistically high value of \logg = 8.6
\citep{2017ASPC..509..195P} would give a spectroscopic distance of
only 140\,pc.

\begin{figure}
\centering
\includegraphics[width=0.9\columnwidth]{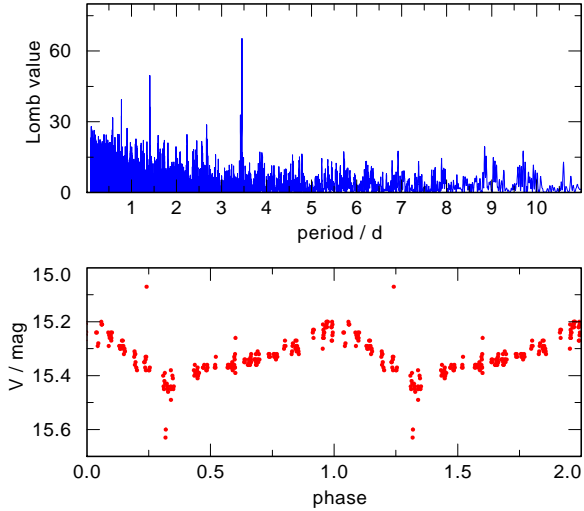}
  \caption{Periodogram (top) and light curve (bottom) of \pg.}
\label{fig:lomb}
\end{figure}

\section{Variability}
\label{sect:variability}

We obtained the V-band light curve of \pg\ from the 2nd data release
of the Catalina Sky Survey and calculated a Lomb-Scargle
periodogram. We found a period of $P = 3.45$\,d (false-alarm
probability $< 10^{-28}$) and an amplitude of 0.19\,mag. Periodogram
and phase folded light curve are shown in Fig.\,\ref{fig:lomb}. The
light curve was classified as of RS CVn type
\citep{2014ApJS..213....9D}. That are usually close binaries with
late-type components, $P$ being the orbital period. We suspect that in
the case of \pg\ a weak magnetic field causes a spotted surface (plus
possible effects of a magnetosphere, see discussion below) and $P$ is
the WD rotation period. 

\section{Results and conclusion}
\label{sect:results}

We have analyzed the UV spectrum of the hot DA \pg. We found that its
effective temperature is significantly lower than recent estimates
from Balmer- and Lyman-line fits. This corroborates earlier results
from detailed UV spectroscopy that the temperature estimates
particularly from Balmer-line fits of the hottest hydrogen-rich WDs
must be regarded as rather uncertain. While \citet{2012zieglerdiss}
found that \Teff\ was often underestimated, we encountered here a case
where \Teff\ was significantly overestimated, so there is no clear
systematic trend. \pg\ was thought to be among the three by far
hottest DA WDs (with \logg $\geq$ 7.0) having \Teff $\approx$
140\,000\,K. It could be that the two others are also cooler, because
their temperature was also derived from Balmer lines. UV spectra are
unavailable for them. The fact that WeDe~1 is not a DAO but a DA
argues against an effective temperature as high as 141\,000\,K (see
below). Our result aggravates the problem that many more non-DA WDs
than DAs are known at the very hot end of the WD cooling sequence. 

The large scatter of results from Balmer-line fits for \pg\ in
different works shows that systematic errors in the data reduction and
fitting techniques are severe in the case of the hottest DAs. The same
holds for the Lyman lines where effects of gravity and temperature are
difficult to disentangle. An important step forward would be a
systematic analysis of available UV spectra of all hot DAs in the
manner presented here.

A further complication of optical spectral analyses is introduced by
the Balmer-line problem, which is encountered with the majority of hot
DAs \citep{2011ApJ...730..128T} and in \pg\ as well
\citep{1994ApJ...432..305B}. It is not entirely removed by using
metal-line blanketed models, as we have seen in our analysis but also,
e.g., in a recent analysis of the DAO EGB\,6, which has similar
atmospheric parameters \citep{2018A&A...616A..73W}. Obviously, there
is a remaining deficit in the model atmospheres. It has been
speculated that the Balmer-line problem might be an indication for the
presence of a \emph{magnetosphere} as recently revealed for a DO WD
exhibiting ultrahigh-ionization lines
\citep{doi:10.1093/mnrasl/sly191}.  If true, then an early speculation
by \citet{1992LNP...401..310N} about the presence of a weak
\emph{photospheric magnetic field} affecting the Balmer lines would
turn out to be partly correct, albeit in a somewhat different sense. A
relation of the Balmer-line problem in \pg\ with the possible presence
of a magnetosphere is corroborated by the fact that the WD exhibits a
periodic variability ($P=3.45$\,d, V-band amplitude 0.19\,mag). This
is similar to what is detected at the above-mentioned DO, which has a
shorter rotation period (0.24\,d) but a similar V-band amplitude
(0.17\,mag). Starspots originating from a magnetic field and/or
geometrical effects of the optically thick, circumstellar
magnetosphere are thought to cause the variability
\citep{doi:10.1093/mnrasl/sly191}. Phase-resolved UV spectroscopy
could shed more light on the origin of the observed optical
variability.

\begin{figure}
 \centering\includegraphics[width=0.9\columnwidth]{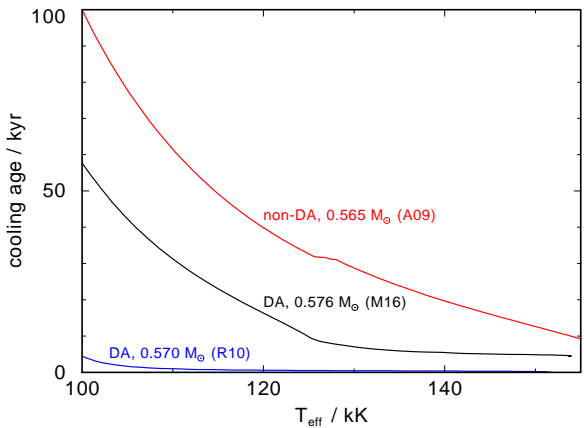}
  \caption{Cooling age of WD models with similar mass (see labels)
    down to \Teff = 100\,000\,K. The non-DA model is a very-late
    thermal pulse track from \citet{2009ApJ...704.1605A} (A09). The
    two DA models are from \citet{2016A&A...588A..25M} (M16) and from
    \citet{2010ApJ...717..183R} (R10). }
\label{fig:cooling}
\end{figure}

Let us look at the cooling times of DA and non-DA WDs in the
high-\Teff\ regime considered in this paper. In
Fig.\,\ref{fig:cooling} we compare the cooling age of three WD models
which have very similar mass, namely one non-DA and two DA models. The
DAs cool faster than the non-DA, but it is surprising that the two DA
models have so different cooling rates. One possible explanation is
that the rate could depend on the thermal-pulse phase when the model
has left the Asymptotic Giant Branch (Miller-Bertolami, priv. comm.;
but note that this is only relevant in the very earliest cooling
phases). Given this uncertainty, it is not possible at the moment to
determine the expected DA/non-DA ratio from the evolutionary rate of
WD models, however, we can make the following rough
estimate. Comparing the DA model of \citet{2010ApJ...717..183R}
with the non-DA model of \citet{2009ApJ...704.1605A}, we notice that
the non-DA needs a factor of 23 more time to cool down to \Teff =
100\,000\,K (and even a factor of 68 to reach \Teff = 120\,000\,K). If
the total DA/non-DA ratio is 5 (as observed from cooler WDs), then the
expected ratio for the very hottest DAs considering a factor 23 in
cooling age is of the order 5/23 $\approx$ 0.22, i.e., it agrees with
the observed DA/non-DA ratio of 0.2 (see Introduction) in the \Teff
$\geq$ 100\,000\,K regime. So it could well be that the paucity of
very hot DAs just reflects their fast cooling rate. The very fast rate
at the highest temperatures (the mentioned factor 68 relative to the
non-DA model) also could explain why the hottest observed non-DAs
reach \Teff = 250\,000\,K, while the hottest DAs have \Teff $\approx$
140\,000\,K.

Finally we notice, as already emphasized, that temperature and gravity
of the DA investigated here (105\,000\,K, \logg = 7.7) are similar to
the DAO EGB\,6 \citep[105\,000\,K, \logg = 7.4,
][]{2018A&A...616A..73W}. EGB\,6 has solar H, He, and metal
abundances, while the He-deficiency and deviations of the metal
abundance pattern from the Solar values indicate that atomic diffusion
is effective in \pg\ because of the slightly higher gravity.



\section*{Acknowledgements}
We thank Pierre Bergeron for sending us his optical spectrum of \pg.
NR is supported by a Royal Commission 1851 research fellowship. The
TMAD tool (\url{http://astro.uni-tuebingen.de/~TMAD}) used for this
paper was constructed as part of the activities of the German
Astrophysical Virtual Observatory. Some of the data presented in this
paper were obtained from the Mikulski Archive for Space Telescopes
(MAST). STScI is operated by the Association of Universities for
Research in Astronomy, Inc., under NASA contract NAS5-26555. Support
for MAST for non-HST data was provided by the NASA Office of Space
Science via grant NNX09AF08G and by other grants and contracts. This
research has made use of NASA's Astrophysics Data System and the
SIMBAD database, operated at CDS, Strasbourg, France. This work has
made use of data from the European Space Agency (ESA) mission {\it
  Gaia} (\url{https://www.cosmos.esa.int/gaia}), processed by the {\it
  Gaia} Data Processing and Analysis Consortium (DPAC,
\url{https://www.cosmos.esa.int/web/gaia/dpac/consortium}). Funding
for the DPAC has been provided by national institutions, in particular
the institutions participating in the {\it Gaia} Multilateral
Agreement.




\bibliographystyle{mnras}
\bibliography{mnras} 




\bsp	
\label{lastpage}
\end{document}